\documentclass[aps,prl,preprint,groupedaddress]{revtex4-1}

\usepackage{amsmath}
\usepackage{amssymb}
\usepackage{graphicx}
\usepackage{mathrsfs}
\usepackage{lineno}
\usepackage{bbm}
\usepackage{mathbbol}
\usepackage{tabularx} % tabularx for automatic full width table
\usepackage{multirow} % multirow for cells that span several rows

\begin{document}
\raggedbottom

\title{Exact analytic expressions for discrete first-passage time probability distributions in Markov networks}

\author{Jaroslav Albert}
%\affiliation{}
%\email{jaroslavalbert81@gmail.com}

\begin{abstract}
The first-passage time (FPT) is the time it takes a system variable to cross a given boundary for the first time. In the context of Markov networks, the FPT is the time a random walker takes to reach a particular node (target) by hopping from
one node to another. If the walker pauses at each node for a period of time drawn from a continuous distribution, the FPT will be a continuous variable; if the pauses last exactly one unit of time, the FPT will be discrete and equal to the number of hops.
We derive an exact analytical expression for the discrete first-passage time (DFPT) in Markov networks. Our approach is as follows: first, we divide each edge (connection between two nodes) of the network
into $h$ unidirectional edges connecting a cascade of $h$ fictitious nodes and compute the continuous FPT (CFPT). Second, we set the transition rates along the edges to $h$, and show that as $h\to\infty$,
the distribution of travel times between any two nodes of the original network approaches a delta function centered at 1, which is equivalent to pauses lasting 1 unit of time. Using this approach, we also compute the joint-probability
distributions for the DFPT, the target node, and the node from which the target node was reached. A comparison with simulation confirms the validity of our approach.
\end{abstract}

\maketitle

\section{Introduction}

First-passage time (FPT) is the time it takes a stochastic system, starting from some initial state $S_i$, to reach some final state $S_f$ for the first time. In a network setting, the system might begin
at a specific node and then, through random jumps, find its way to the target node or nodes. The jumps may be continuous or discrete in time. In the former case the ``walker" would pause at each node for 
a time sampled from a continuous distribution, which gives rise to a continuous FPT (CFPT) distribution. In the latter case, the pauses last exactly 1 unit of time, e. g. 1 second, every time, which results in a discrete FPT (DFPT) distribution.
The FPT is generally difficult to compute analytically, which is why the literature offers plenty of papers that 
focus on the mean first-passage time, both continuous \cite{Condamin,Lau,Zhang,Foster,Ma} and discrete \cite{Chen,Huang}, while lacking a simple formula for the FPT distributions for complex networks.

In this paper we present a simple and straight-forward way to compute the exact DFPT distributions for a random walk on a network. Instead of focusing on the target nodes, we instead work with the edges (connections),
i. e. transition steps between nodes, that lead to the target nodes. Applying a theorem from our previous work \cite{Albert}, we show how a simple modification of the Master equation truns into an equation for the CFPT distribution.
To model the discrete time transitions between nodes, we turn each edge into a cascade of $h$ fictitious transitions, each with a transition rate $h$, and show that in the limit as $h\to\infty$ the travel time between nodes approaches a delta function centered at 1. The outcome of this scheme
is a solution for the DFPT that is exact, simple and easy to compute for large networks. Furthermore, this approach allows one also to compute the joint-probability $P_q(k\to p)$ that the final transition occurred at a discrete time $q$ and between the
node $k$ and the target node $p$.

\section{Continuous FPT distributions}

The Master equation for a network of $N$ nodes is given by
\begin{equation}
\frac{d{\bf P}}{dt}={\bf T}{\bf P},
\end{equation}
whith ${\bf P}$ being the state probability vector and ${\bf T}$ the transition matrix
\begin{eqnarray}\label{ToffD}
	T_{ij}=\left[\begin{array}{cccccccccc}
		-{\bar T}_1 & T_{12}      & \,\,\,\cdot\,\,\, & \,\,\,\cdot\,\,\, & \,\,\,\cdot\,\,\, & \,\,T_{1N}  \\
		
		     T_{21} & -{\bar T}_2 & \cdot & \cdot & \cdot & T_{2N}  \\
	
		        \cdot &                 & \cdot &          &         &  \cdot             \\
		
		        \cdot &                 &         & \cdot &          &  \cdot    \\

		        \cdot &                 &         &         & \cdot  &  \cdot    \\
		
		        T_{N1} & T_{N2}  & \cdot & \cdot & \cdot & -{\bar T}_N \\
	\end{array}\right],
\end{eqnarray}
where ${\bar T}_i=\sum_jT_{ij}$.
\vspace{1cm}

{\it Theorem:}

For arbitrary initial conditions
${\bf P}(0)$, the FPT probability distribution, $P_{\text{FPT}}(t)$, is given by
\begin{equation}\label{PFPT}
P_{\text{FPT}}(t)=-\frac{d}{dt}\sum_{j=1}^nQ_j(t),
\end{equation}
such that ${\bf Q}(t)=[Q_1(t),...,Q_N(t)]$ satisfies
\begin{equation}\label{Qvector}
\frac{d{\bf Q}}{dt}=[{\bf T}'-{\bf R}]{\bf Q},\,\,\,\,\,\,{\bf Q}(0)={\bf P}(0),
\end{equation}
where
${\bf T}'=T_{ij}(1-\delta_{ij})$ if both $i$ and $j$ are non-target nodes, otherwise $T'_{ij}=0$; and
${\bf R}=\delta_{ij}({\tilde T}_l+\sum_lT_{il})$ ($l=$ all non-target nodes) is a diagonal matrix. 
\vspace{1cm}

\noindent
For proof refer to \cite{Albert}.

Although the above theorem includes cases where the initial node is one of the target nodes, in this paper we chose to focus on those instances where the initial node is non-target (see Fig. 1).
In such a class of problems, it is convenient to relabel the nodes so that the target nodes are $[n+1,...,N]$ and the non-target nodes are $[1,...,n]$. This set the dimension of the matrices ${\bf T}'$ and ${\bf R}$
to $n$:
\begin{eqnarray}\label{ToffD}
	T^{'}_{ij}=\left[\begin{array}{cccccccccc}
		0 & T_{12}      & \,\,\,\cdot\,\,\, & \,\,\,\cdot\,\,\, & \,\,\,\cdot\,\,\, & \,\,T_{1n}  \\
		
		     T_{21} & 0 & \cdot & \cdot & \cdot & T_{2n}  \\
	
		        \cdot &                 & \cdot &          &         &  \cdot             \\
		
		        \cdot &                 &         & \cdot &          &  \cdot    \\

		        \cdot &                 &         &         & \cdot  &  \cdot    \\
		
		        T_{n1} & T_{n2}  & \cdot & \cdot & \cdot & 0 \\
	\end{array}\right],\,\,\,\,\,
R_{ij}=\left[\begin{array}{cccccccccc}
		         -r_1 & 0      & \,\,\,\cdot\,\,\, & \,\,\,\cdot\,\,\, & \,\,\,\cdot\,\,\, & 0  \\
		
		             0 & -r_2 & \cdot & \cdot & \cdot & 0  \\
	
		        \cdot &                 & \cdot &          &         &  \cdot             \\
		
		        \cdot &                 &         & \cdot &          &  \cdot    \\

		        \cdot &                 &         &         & \cdot  &  \cdot    \\
		
		             0 & 0  & \cdot & \cdot & \cdot & -r_n    \\
	\end{array}\right],
\end{eqnarray}
where
\begin{equation}\label{Rmatrix}
r_i={\tilde T}_i+\sum_{k=n+1}^NT_{ki}.
\end{equation}
\begin{figure}\label{Fig1}
	\centering
	\includegraphics[trim=0 0 0 1.0cm, height=0.3\textheight]{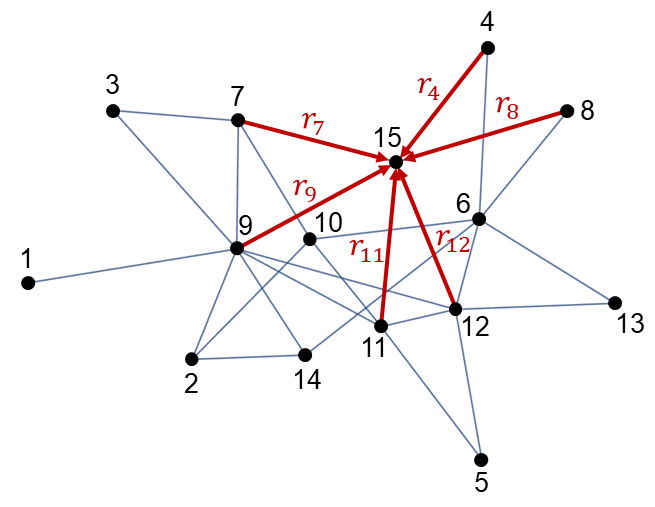}
	\caption{An example of a random, undirected network with one target node (node 15). The red arrows show all the transitions leading to the target node, as well as the non-zero elements of the matrix ${\bf R}$ (see Eq. (\ref{Rmatrix})).}
\end{figure}

\section{Discrete FPT distributions}

We can use the above theorem to derive exact expressions for the DFPT distribution by introducing a cascade of transitions along each edge, as shown in Figure 2. 
As we are about to show, when the number $h$ and transition rate $a$ of these fictitious transitions becomes infinite, such that $a=h\to\infty$, the time it takes to traverse one edge becomes 1.
To prove this, we need only to insert the appropriate matrices ${\bf T}'$ and ${\bf R}$ into Eq. (\ref{Qvector}) and solve it for ${\bf Q}$ with the initial condition ${\bf Q}(0)={\bf P}(0)=\delta_{i1}$.

\begin{figure}\label{Fig2}
	\centering
	\includegraphics[trim=0 0 0 1.0cm, height=0.3\textheight]{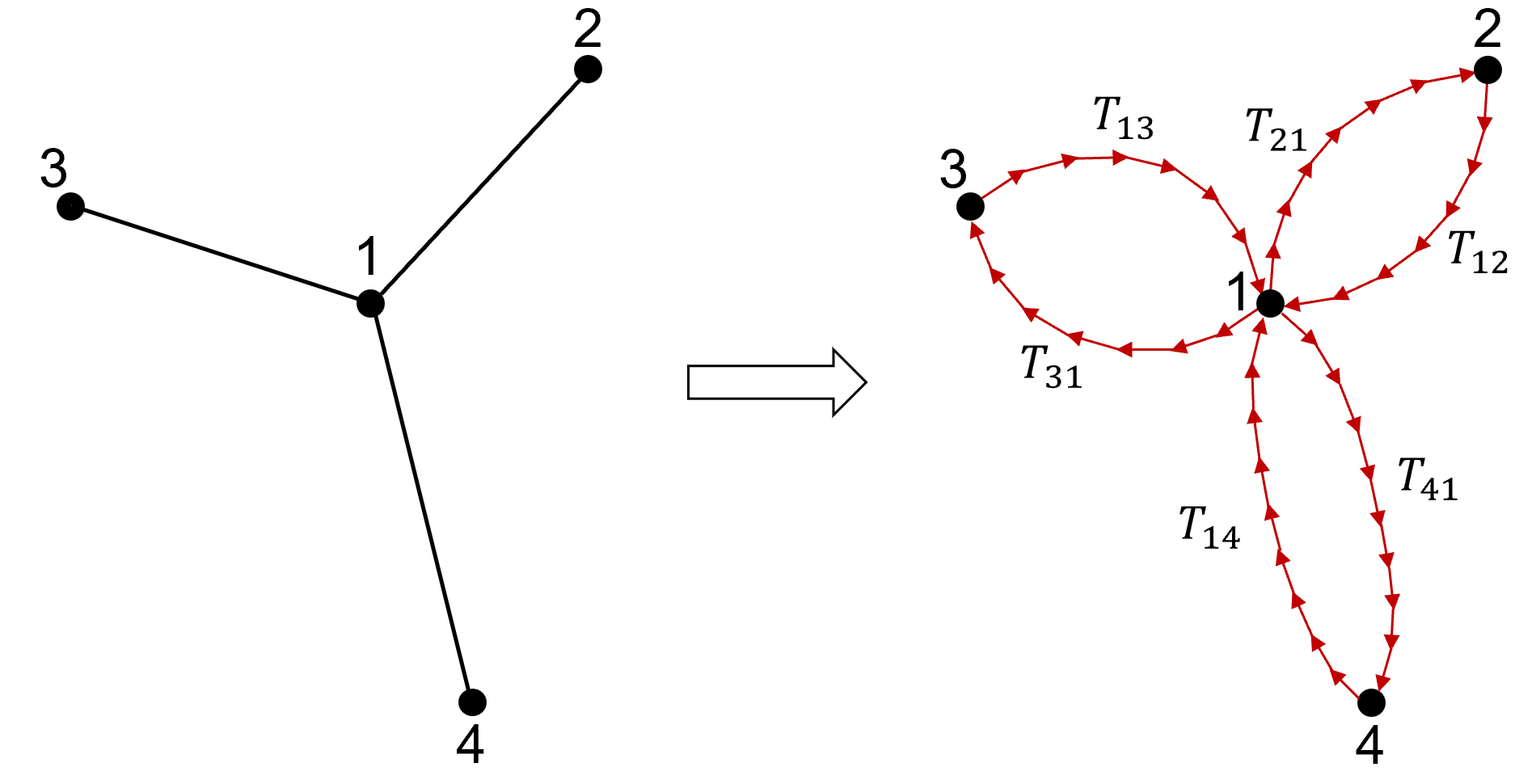}
	\caption{Each node of a network (left) is replaced by a cascade of fictitious transitions (right). If the network is directed, each edge has only one cascade; for undirected networks, each direction,
i. e. $i\to j$ and $j\to i$, must have its own cascade.}
\end{figure}

%\begin{figure}\label{Fig1}
%	\centering
%	\includegraphics[trim=0 0 0 1.0cm, height=0.06\textheight]{Cascade}
%	\caption{...}
%\end{figure}

Taking a Laplace transform of Eq. (\ref{Qvector}), the set of desired equations become:
\begin{eqnarray}
sQ_1&=&1-aQ_1,\nonumber\\
sQ_2&=&aQ_1-aQ_2,\nonumber\\
&\cdot&\nonumber\\
&\cdot&\nonumber\\
&\cdot&\nonumber\\
sQ_{h-1}&=&aQ_{h-2}-aQ_{h-1}\nonumber\\
sQ_{h}&=&aQ_{h-1}-2aQ_{h}.
\end{eqnarray}
The solution of $Q_i$ is
\begin{equation}
Q_p=\frac{1}{a}\left(\frac{a}{a+s}\right)^{p},\,\,\,\,\,\,\,\text{for}\,\,\, 0<p<h,\,\,\,\,\,\,\,\,Q_h=\frac{1}{2a+s}\left(\frac{a}{a+s}\right)^{h}
\end{equation}
which yields
\begin{equation}
Q(s)=\sum_{p=1}^{h}Q_p=\frac{1}{s}\left[1-\left(\frac{a}{a+s}\right)\right]-\frac{1}{a}\left(\frac{a}{a+s}\right)+\frac{1}{2a+s}\left(\frac{a}{a+s}\right)^{h}
\end{equation}
Taking the limit $a=h\to\infty$, we obtain $Q(s)=(1-e^{-s})/s$, of which the inverse Laplace transform is $Q(t)=\theta(t)-\theta(t-1)$. The first term simply means that $t$ must be strictly $>0$.
Plugging $Q(t)$ into Eq. (\ref{PFPT}) gives the PFT distribution $\delta(t-1)$ for $t>0$.

Let us now write Eq. (\ref{Qvector}) for a network of $N$ nodes, with each node being a cascade of $h$ transitions.  
\begin{eqnarray}
&&\frac{dQ_{ij}^{(m)}}{dt}=h\left[T'_{ij}\delta_{m1}Q_j+(1-\delta_{m1})Q_{ij}^{(m-1)}-Q_{ij}^{(m)}\right],\label{Qmij}\,\,\,\,\,\,\,\,0<m\leq h\\
&&\frac{dQ_i}{dt}=h\sum_j\left[Q_{ij}^{(h)}-r_j\delta_{ij}Q_j\right].\label{Qj}
\end{eqnarray}
The index $m$ in $Q_{ij}^{(m)}$ gives the location along the cascade on the edge $(i,j)$, while the index $j$ in $Q_j$ refers to node $j$.
Before we continue, there is a discrepancy that needs to be pointed out. The first-passage time reactions we have chosen, and that appear in the diagonal matrix elements $r_j$, are not the ones that lead to the target nodes, but
rather those that go from a node $j$ to the cascade position $m=1$ along the edge $(j,k)$, where $k$ is a target node. As a result, the solution to Eqs. (\ref{Qmij}) and (\ref{Qj}) will correspond to the DFPT variable that is smaller than the actual DFPT variable by 1.
To fix this discrepancy, all we need to do is shift the time variable by $-1$: $Q(t)\to Q(t-1)$. 
With this in mind, we can continue as we were.
Taking the Laplace transform of Eqs. (\ref{Qmij}) and (\ref{Qj}), and rearranging some terms, leads to
\begin{eqnarray}
&&Q_{ij}^{(m)}=\frac{h}{s+h}\left[(1-\delta_{m1})Q_{ij}^{(m-1)}+T'_{ij}\delta_{m1}Q_j\right],\label{LQmij}\\
&&-P_i(0)+sQ_i=h\sum_j\left[Q_{ij}^{(h)}-r_j\delta_{ij}Q_j\right],\label{LQj}
\end{eqnarray}
where we chose the initial conditions to be $Q^{(m)}_{ij}(0)=0$ and $Q_j(0)=\delta_{j1}$, since we must start on a real node.
Equation (\ref{LQmij}) can be solved in terms of $Q_j$:
\begin{equation}\label{LQmij3}
Q_{ij}^{(m)}=T'_{ij}\left[\frac{h}{s+h}\right]^mQ_j.
\end{equation}
Defining a new variable ${\tilde Q}_i=hQ_i$, Eq. (\ref{LQj}) can now be expressed as
\begin{equation}
-P_i(0)+\left(\frac{s}{h}\right){\tilde Q}_i=\sum_j\left[\left(\frac{h}{s+h}\right)^hT'_{ij}-r_j\delta_{ij}\right]{\tilde Q}_j.
\end{equation}
Taking the limit $h\to\infty$, we obtain
\begin{equation}\label{LQj2}
\sum_j\left[e^{-s}T'_{ij}-r_j\delta_{ij}\right]{\tilde Q}_j=-P_i(0).
\end{equation}
For convenience we now switch to vector notation and write, ${\bf\tilde Q}=({\tilde Q}_1,...,{\tilde Q}_n)$, 
${\bf T}'=T'_{ij}$, ${\bf R}=r_j\delta_{ij}$ and 
${\bf P}(0)=(P_1(0),...,P_n(0))$, which casts Eq. (\ref{LQj2}) into the form:
\begin{equation}
{\bf T}'{\bf\tilde Q}e^{-s}-{\bf R}{\bf\tilde Q}=-{\bf P}(0)
\end{equation}
Expressing the solution in powers of $e^{-s}$, i. e.
\begin{equation}
{\bf\tilde Q}=\sum_{q=0}{\bf f}_qe^{-qs},
\end{equation}
we obtain
\begin{equation}
\sum_{q=0}[{\bf T}'{\bf f}_qe^{-s(q+1)}-{\bf R}{\bf f}_qe^{-sq}]=-\sum_{q=0}{\bf P}(0)\delta_{q0}e^{-qs}.
\end{equation}
Equating coefficients with like exponentials leads to an equation for ${\bf f}_{q}$
\begin{equation}
{\bf R}{\bf f}_{q}={\bf T}'{\bf f}_{q-1}+{\bf P}(0)\delta_{q0},
\end{equation}
the solution of which is
\begin{equation}\label{fq}
{\bf f}_{q}=\left({\bf R}^{-1}{\bf T}'\right)^q\left({\bf R}^{-1}{\bf P}(0)\right).
\end{equation}
We need to compute
\begin{equation}\label{Qs}
Q(s)=\sum_{ijm}Q_{ij}^{(m)}(s)+\sum_jQ_j(s).
\end{equation}
Invoking Eq. (\ref{LQmij3}), the first term on the right hand side can be written as
\begin{equation}
\sum_{ij}\sum_{m=1}^hQ_{ij}^{(m)}(s)=\sum_{ij}T'_{ij}\frac{{\tilde Q}_j}{h}\sum_{m=1}^h\left[\frac{h}{s+h}\right]^m.
\end{equation}
By virtue of the identity
\begin{equation}
\sum_{m=1}^h\left[\frac{h}{s+h}\right]^m=\frac{h}{s}\left[1-\left(\frac{h}{h+s}\right)\right],
\end{equation}
and the limit $h\to\infty$, we can write
\begin{equation}\label{a25}
\sum_{ij}\sum_mQ_{ij}^{(m)}(s)=\sum_{ij}T'_{ij}{\tilde Q}_j\left[\frac{1-e^{-s}}{s}\right].
\end{equation}
In that same limit, the second term on the right hand side of Eq. (\ref{Qs}) goes to zero, since $Q_j={\tilde Q}_j/h\to 0$ as $h\to\infty$.
Thus, inserting Eq. (\ref{fq}) into Eq. (\ref{a25}) yields
\begin{equation}
Q(s)=\sum_{q=0}\Lambda_q\left[\frac{e^{-qs}-e^{-s(q+1)}}{s}\right]=\sum_{q=0}(\Lambda_q-\Lambda_{q-1})\frac{e^{-qs}}{s}
\end{equation}
where
\begin{equation}\label{Lambda}
\Lambda_q=\sum_{ij}T'_{ij}({\bf f}_q\cdot{\bf u}_j),
\end{equation}
where ${\bf u}_j$ is the $j^{\text{th}}$ unit vector of dimension $n$.
The Laplace transform of $P_{\text{FPT}}(t)$, as defined in Eq. (\ref{PFPT}), is
\begin{equation}\label{LapP}
P_{\text{FPT}}(s)=-\int_0^\infty dt\frac{dQ(t)}{dt}=Q(t=0)-sQ(s)=1-\sum_{q=0}(\Lambda_q-\Lambda_{q-1})e^{-qs},
\end{equation}
where $P_{\text{FPT}}(0)=P_{\text{FPT}}(t=0)$.
Taking the inverse Laplace transform of Eq. (\ref{LapP}), we obtain
\begin{equation}
P_{\text{FPT}}(t)=(1-\Lambda_0)\delta(t)+\sum_{q=1}(\Lambda_{q-1}-\Lambda_q)\delta(t-q).
\end{equation}
Remembering to shift $t$ by -1, the FPT distribution reads:
\begin{equation}
P_{\text{FPT}}(t)=(1-\Lambda_0)\delta(t-1)+\sum_{q=1}(\Lambda_{q-1}-\Lambda_q)\delta(t-q-1),
\end{equation}
So, the discrete probability that any one of the states $n+1,...,N$ is reached from state 1 in $q$ steps is given by
\begin{equation}\label{Pq}
P_q=(1-\Lambda_0)\delta_{q1}+[\Lambda_{q-2}-\Lambda_{q-1}](1-\delta_{q1}).
\end{equation}

\section{Joint probability distributions}
\begin{figure}\label{Fig3}
	\centering
	\includegraphics[trim=0 0 0 1.0cm, height=0.337\textheight]{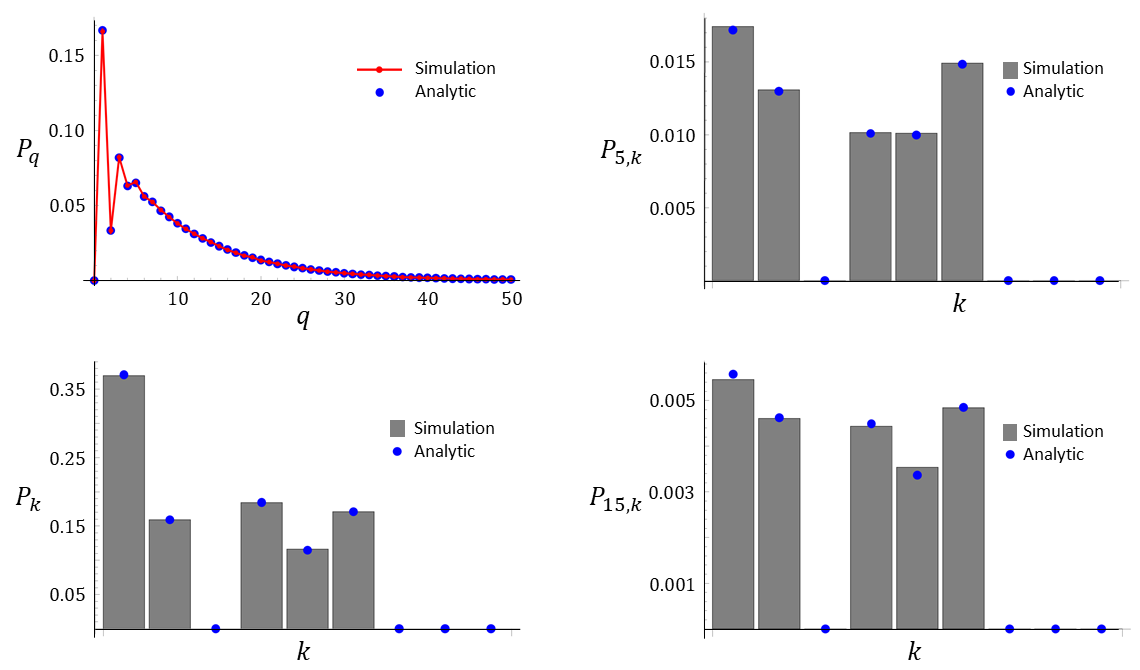}
	\caption{Comparison between simulation and analytic expressions for $P_q$, $P(k)$, $P_5(k\to N)$ and $P_{10}(k\to N)$ for a random undirected network with $N=10$ nodes
and network density 0.5. The ensemble size for the simulation was 200k.}
\end{figure}
\begin{figure}\label{Fig4}
	\centering
	\includegraphics[trim=0 0 0 1.0cm, height=0.9\textheight]{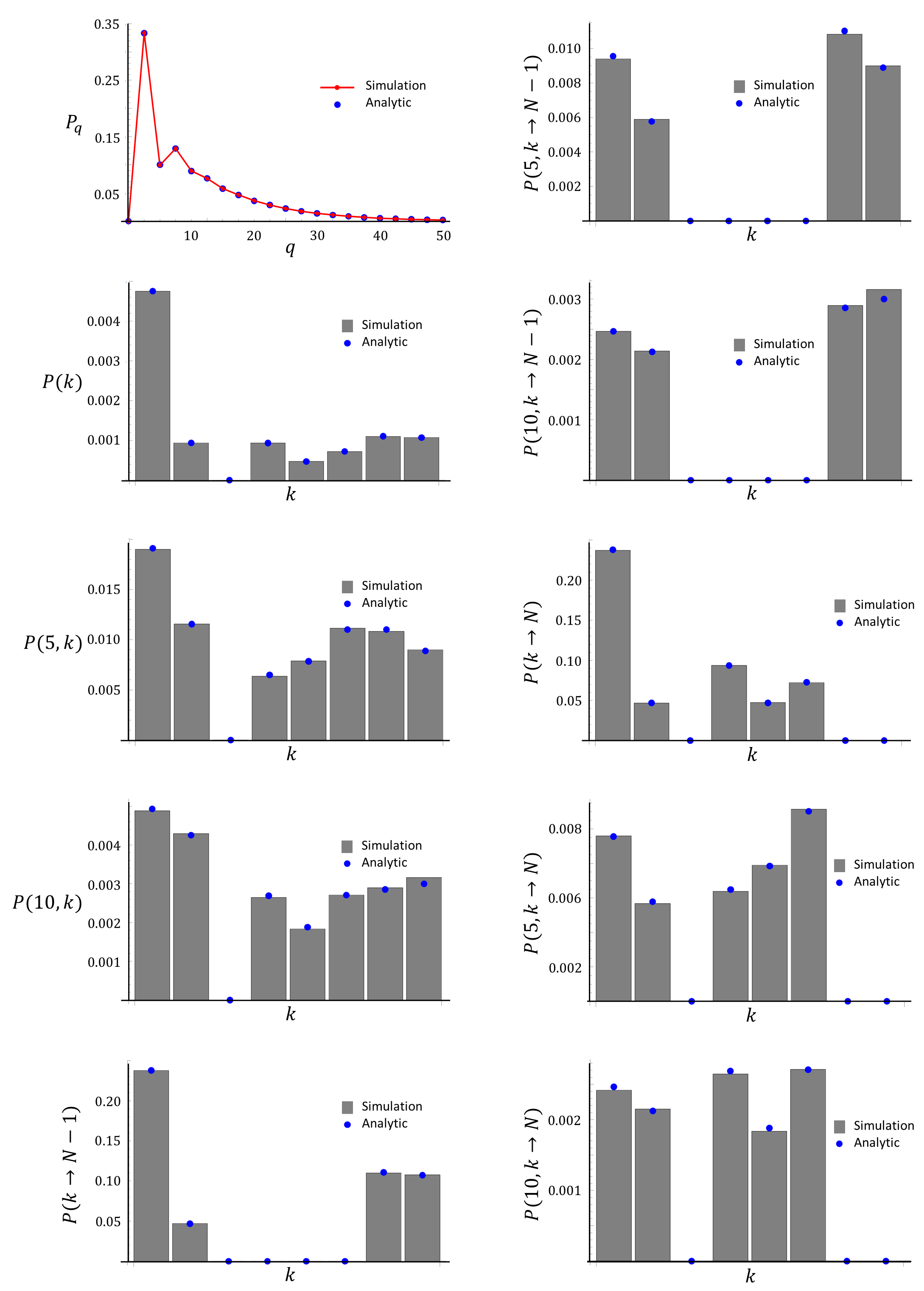}
	\caption{Same as in Fig. 3 but with two target nodes: $N-1$ and $N$.}
\end{figure}
We can also compute the joint probability distribution, $P_{\text{FPT}}(t,k\to p)$, for the FPT and the reaction $k\to p$, where $p$ is a target node \cite{Albert2}.
\begin{equation}\label{TheResult}
P_{\text{FPT}}(t,k\to p)=Q_k(t-1)hT_{pk}={\bf\tilde Q}\cdot{\bf u}_kT_{pk}=\sum_{q=0}T_{pk}({\bf f}_q\cdot{\bf u}_k)\delta(t-q-1).
\end{equation}
The coefficients give the joint-probability for DFPT, the target node $p$, and the node $k$ from which $p$ was reached.
\begin{equation}\label{Pqjk}
P_q(k\to p)=T_{pk}({\bf f}_{q-1}\cdot{\bf u}_k).
\end{equation}
From this equation, we can recover Eq. (\ref{Pq}) by summing over $k$ and $p$,
\begin{equation}\label{Pqrecovered}
P_q=\sum_{pk}T_{pk}({\bf f}_{q-1}\cdot{\bf u}_k)=(1-\Lambda_0)\delta_{q1}+[\Lambda_{q-2}-\Lambda_{q-1}](1-\delta_{q1}),
\end{equation}
(see the Appendix) or the probability that a target node was reached from the node $k$ regardless of the FPT by summing over $q$ and $p$:
\begin{equation}\label{Pkkk}
P(k)=\sum_{q=0}^{\infty}\sum_pT_{pk}({\bf f}_{q-1}\cdot{\bf u}_k).
\end{equation}
Figures 3 and 4 show a comparison between simulation and Eqs. (\ref{Pq}), (\ref{Pqjk}) and (\ref{Pkkk}) for a single node target and a two-node target, respectively.

\section{Conclusion}

We have derived exact analytic expressions for the discrete first-passage time (DFPT) in Markov networks. By introducing a cascade of $h$ fictitious transitions along all
edges, each with a transition rate $h$, we showed that in the limit as $h\to\infty$, the continuous FPT (CFPT) distribution approaches the DFPT. Using this approach, we also computed the
joint probability distribution for the DFPT, the target node, and the node from which the target node was reached.
Computation of the joint-probability requires $q-1$ number of matrix multiplications, where $q$ is the DFPT variable, while the DFPT probability also involves a sum over the number of edges.
Thus, the time complexity of our method is at most ${\cal O}(qEn^{2.371552})$, where $E$ is the number of edges. Given its simplicity and speed, our method can be extended to other network-like systems,
such as particles diffusing on a lattice, and Brownian dynamics.

%Although $P_q$ Eq. (\ref{Pq}) is analytic, it contains the matrix ${\bf M}^{-1}{\bf T}'$ raised to some power $q$ (Eq. (\ref{fq})), as well as a double summation over
%the nodes (Eq. (\ref{Lambda})). It is therefore of interest to estimatite the time complexity for computing $P_q$.
%TC (time complexity) of finding eigenvalues: $N^3$.
%TC of finding eigenvectors: $N^2$.
%TC for multiplying matrices: $n^{2.371552}$.
%
%If the range of $q$ is small, than use matrix multiplication; otherwise use eigenvalues.

%Show that $\sum_{jk}P_q(j\to k)=P_q$.

%\frac{k^2}{N}e^{-k^2/N}+\sqrt{2\pi k}\left(\frac{k}{N}\right)^k\text{exp}\left[\left(\frac{1}{2N}+\alpha\right)k^2-(1+\alpha)k\right]

%One of the most famous random graph is the Erdös-Rényi model G(n, p)
%with n vertices and each edge is added with probability p.

\end{document}